\documentclass{article}

\usepackage{PRIMEarxiv}

\usepackage[utf8]{inputenc} 
\usepackage[T1]{fontenc}    
\usepackage{hyperref}       
\usepackage{url}            
\usepackage{booktabs}       
\usepackage{amsfonts}       
\usepackage{nicefrac}       
\usepackage{microtype}      
\usepackage{lipsum}
\usepackage{fancyhdr}       
\usepackage{graphicx}       
\graphicspath{{media/}}     
\usepackage{amsmath, amssymb, amsfonts}
\usepackage{geometry}
\usepackage{qcircuit}
\usepackage{microtype}

\title{Quantum Circuit and Tensor Network Implementation of the 2D Acoustic Wave Equation}

\author{Tamas Nemeth \\
   WaveRunr\\
   \texttt{tamas.nemeth@waverunr.com}\\
   \And
  Gábor Vattay \\
  Institute of Physics and Astronomy\\
  Eötvös Loránd University\\
  Budapest, Hungary\\
  \texttt{gabor.vattay@ttk.elte.hu} \\
}

\newcommand{\ket}[1]{|#1\rangle}
\newcommand{\bra}[1]{\langle#1|}

\begin{document}


\renewcommand{\thefootnote}{\fnsymbol{footnote}}

\maketitle

\begin{abstract}
  We present a cohesive framework for simulating seismic wave propagation utilizing quantum computing paradigms and their classical tensor network equivalents. 
  We detail a quantum circuit-based formulation for the explicit finite-difference time-domain (FDTD) solution of the two-dimensional acoustic wave equation and
  map this quantum architecture onto a tensor train representation, namely for Matrix Product State (MPS).
  The MPS solver enables deterministic simulation of large-scale wavefield dynamics on classical high-performance computing systems.
  We demonstrate the MPS representation by computing 2D seismic wavefields on the Marmousi model.
  Our results indicate that the MPS representation is a viable direction for computing and scaling wavefield propagation.
\end{abstract}

\section{Introduction}

The accurate simulation of seismic wave propagation through complex subsurface media is a foundational task in reflection seismology, underpinning advanced seismic imaging methods such as Reverse Time Migration (RTM) and Full Waveform Inversion (FWI). 
High-resolution seismic surveys push classical high-performance computing for wavefield propagation to its architectural limits.
Reducing the cost of wavefield propagation is therefore of fundamental interest, 
whether to prioritize more efficient hardware implementations or novel algorithmic frameworks.

Quantum computing offers a new computational and algorithmic paradigm with potential exponential memory scaling inherent to massive multi-dimensional grids such as those used for seismic imaging.
Quantum hardware to utilize these benefits at this scale is not yet available, although quantum technology is rapidly improving.
Quantum algorithms for circuit-based quantum systems, the leading quantum hardware technology among a few others,
enable arbitrarily general computational tasks; and their emulation for classical hardware is also possible using tensor networks. 
\cite{oseledets2010tt} developed a tensor network methodology, called a tensor train, that is especially suitable 
for computational tasks such as wavefield propagation.
Tensor train methods operate using matrix product states (MPS) where the step-wise matrix-matrix multiplies
correspond to classical time steppers.

Recently, several studies have been conducted to evaluate quantum computing methods for wavefield propagation.
\cite{fichtner2024&1Delastic} implemented the 1D discrete elastic wave equation to a Schr\"odinger equation that
can be run on a circuit-based quantum computer.
\cite{bosch2025&qwavesimulation} extended it to a  broader class of wave equations, including the acoustic wave equation, 
Maxwell’s equations, and the elastic wave equation.
Previously, \cite{fomel2013&lowrank} used low-rank approximations for finite-difference seismic wave extrapolation for the 
acoustic wave equation. Their work shares structural similarities with tensor networks. 

In this study we derive the 2D acoustic wave equation using both quantum encoding suitable for quantum circuits and the corresponding version in MPS formulation. 
We implement the MPS solver on 2D Marmousi data and compare it with the explicit finite-difference time-domain (FDTD) solver.
While the explicit derivation of the quantum encoding and MPS solver methods 
is rigorous and comprehensive, the goal is to 
illustrate and compare the resulting algorithms with the FDTD solution.
Crucially, we implement the MPS solver on a 2D model to enhance the practical relevance of these algorithms.


\section{Acoustic wave equation in 2D}

The propagation of a scalar acoustic wavefield $u(x,z,t)$ in an inhomogeneous velocity medium in 2 dimensions is governed by the partial differential equation:  
\begin{equation}
    \frac{\partial^2 u(x,z,t)}{\partial t^2} = c^2(x,z) \left( \frac{\partial^2 u}{\partial x^2} + \frac{\partial^2 u}{\partial z^2} \right),
\end{equation}
where $x$ is the horizontal spatial coordinate, $z$ represents depth, and $c(x,z)$ denotes the spatially varying velocity model of the subsurface.

We discretize the continuous domain onto a square Cartesian grid defined by $N_x = N_z = 2^n$ points, with a uniform spatial grid spacing $\Delta h = \Delta x = \Delta z$. Time is discretized by a temporal step $\Delta t$, strictly constrained by the Courant-Friedrichs-Lewy (CFL) stability criterion. Applying a second-order central difference scheme to both the temporal and spatial derivatives yields the explicit FDTD update equation:
\begin{align}
    u_{i,j}^{k+1} & = 2u_{i,j}^k - u_{i,j}^{k-1} + \nonumber  \\ 
    & + \lambda^2 c_{i,j}^2 \left( u_{i+1,j}^k + u_{i-1,j}^k + u_{i,j+1}^k + u_{i,j-1}^k - 4u_{i,j}^k \right),
\end{align}
where the indices $i$ and $j$ denote the spatial grid coordinates in the $x$ and $z$ directions, respectively, $k$ denotes the discrete time step, and $\lambda=\frac{\Delta t}{\Delta h}$ is the mesh ratio.

\section{Quantum Register Encoding}

To embed this classical FDTD scheme into a quantum processor, the $2^n \times 2^n$ grid is mapped directly onto the computational basis of a quantum register comprising $2n$ qubits. The state of the acoustic wavefield at time step $k$ is represented by the unnormalized quantum state vector:
\begin{equation}
    \ket{\psi^k} = \sum_{x=0}^{2^n-1} \sum_{z=0}^{2^n-1} u_{x,z}^k \ket{x} \ket{z},
\end{equation}
where $\ket{x} = \ket{x_{n-1} \dots x_1 x_0}$ and $\ket{z} = \ket{z_{n-1} \dots z_1 z_0}$ are the binary representations of the spatial indices.

\vspace{\baselineskip}
\vspace{\baselineskip}
The squared velocity field $c_{i,j}^2$ contains the spatially variable geophysical properties of the medium. In the computational basis, this maps to a diagonal operator $\hat{C}^2$ that applies a coordinate-dependent amplitude scaling to the corresponding quantum states:
\begin{equation}
    \hat{C}^2 \ket{x}\ket{z} = c^2(x,z) \ket{x}\ket{z}.
\end{equation}

For a complex geophysical model, $\hat{C}^2$ is implemented via the quantum Read-Only Memory (QROM) architectures in \cite{babbush2018encoding} or approximated through a Walsh series expansion. The physical constraint $c(x,z) > 0$ ensures that the operator is strictly positive definite.

\subsection{Quantum Shift Operators via Arithmetic Circuits}

The application of the discrete spatial Laplacian operator requires translating the amplitude field along the coordinate axes. To mitigate the phase errors and deep routing overhead associated with the quantum Fourier Transform-based translation, we implement the forward and backward shift operators, $\hat{S}^+$ and $\hat{S}^-$, strictly in the computational basis using quantum incrementer and decrementer circuits.

The forward shift operator $\hat{S}_x^+$ performs the cyclic mapping $\ket{x} \to \ket{(x+1) \bmod 2^n}$ as shown in \cite{vedral1996quantum}. This operation is mathematically equivalent to a ripple-carry adder with one input fixed to unity. The transformation on the \mbox{$m$-th} qubit of the register relies on the logical AND of all less significant bits:
\begin{equation}
    x_m \leftarrow x_m \oplus \left( \prod_{l=0}^{m-1} x_l \right).
\end{equation}

In the quantum circuit model, this requires a cascaded sequence of multi-controlled Pauli-X (Toffoli) gates. The circuit depth is minimized by executing the gates sequentially from the most significant bit (MSB) down to the least significant bit (LSB). The quantum circuit for a 3-qubit incrementer ($\hat{S}^+$) is depicted below:

\vspace{1em}
\begin{center}
\centerline{
\Qcircuit @C=1.5em @R=1.5em {
\lstick{\ket{x_0} \text{ (LSB)}} & \ctrl{2} & \ctrl{1} & \gate{X} & \qw \\
\lstick{\ket{x_1}}               & \ctrl{1} & \targ    & \qw      & \qw \\
\lstick{\ket{x_2} \text{ (MSB)}} & \targ    & \qw      & \qw      & \qw
}
}
\end{center}
\vspace{-1.em}
Figure 1: The quantum circuit for a 3-qubit incrementer

\vspace{0.5cm}
The backward shift operator $\hat{S}_x^-$ achieves $\ket{x} \to \ket{(x-1) \bmod 2^n}$. 
Because quantum evolution is unitary, $\hat{S}_x^-$ is exactly the adjoint $(\hat{S}_x^+)^{\dagger}$, implemented by reversing the gate order of the incrementer circuit. The discrete Laplacian operator is consequently formed by the sum of these unitaries acting on their respective registers:
\begin{equation}
    \hat{L} = \hat{S}_x^+ \otimes \hat{I}_z + \hat{S}_x^- \otimes \hat{I}_z + \hat{I}_x \otimes \hat{S}_z^+ + \hat{I}_x \otimes \hat{S}_z^- - 4\hat{I}.
\end{equation}
Compare this equation to the second row of equation (2) to understand the structure of the Laplacian in the traditional FD native domain and using quantum shift operators.

\subsection{Wavefield evolution via Linear Combination of Unitaries}

The explicit FDTD stepping requires the two most recent steps to complete the next step.
The state update is formalized as:
\begin{equation}
    \ket{\psi^{k+1}} = \hat{A}\ket{\psi^k} - \ket{\psi^{k-1}},
\end{equation}
where $\hat{A} = 2\hat{I} + \lambda^2 \hat{C}^2 \hat{L}$ is a generalized non-unitary transition operator. 
We recognize that $\hat{A}$ is the sum of constituent unitary operations scaled by non-unitary coefficients. We deploy the Linear Combination of Unitaries (LCU) protocol to block-encode $\hat{A}$.

To map the second-order temporal stencil, we maintain two working registers for consecutive time steps, $\ket{\psi^k}$ and $\ket{\psi^{k-1}}$. The subtraction operator is implemented utilizing controlled-SWAP operations intertwined with phase inversions, completing the synthesis of the exact numerical wavefield integ-ration onto the quantum architecture.

\section{Motivation for Tensor Train Emulation of Subsurface Models}

Current quantum hardware constraints, specifically the absence of deep fault-tolerance required by the LCU algorithms, call for intermediate classical emulation strategies. The Tensor Train (TT) framework, mathematically isomorphic to MPS methods in quantum many-body physics, provides a rigorous bridge. It exploits the identical spatial entanglement bounding principles as the quantum algorithm but executes deterministically on classical high-performance computing architectures.

In the context of seismic wave propagation, quantum entanglement corresponds directly to the spatial correlation and structural complexity of the wavefield. Propagation through smoothly varying subsurface velocity models preserves a low entanglement signature, rendering the $2^{2n}$-dimensional state vector highly compressible into a low-rank tensor network (\cite{fomel2013&lowrank}). This paradigm facilitates the exact exe-cution of the quantum algorithm's operational logic while managing the exponential memory limitations inherent to classical von Neumann architectures such CPUs or GPUs.

\subsection{Tensor Network Representation}

To circumvent the exponential memory requirements of scaling multidimensional grids, the $N_x \times N_z$ discrete field is mapped onto a so-called many-body quantum spin chain. Because $N_x = N_z = 2^n$, the total number of grid points is $2^{2n}$. The state of the field is isomorphic to a state vector in a $2n$-qubit Hilbert space $\mathcal{H} \cong (\mathbb{C}^2)^{\otimes 2n}$.

The discrete field amplitudes $u_{i,j}$ are reshaped into a rank-$2n$ tensor, which is subsequently decomposed into a matrix product state via sequential Singular Value Decompositions (SVD). The state is expressed equivalently to the quantum register formulation as an open tensor product shown in \cite{schollwock2011density}:
\begin{equation}
    \ket{\psi^k} = \sum_{s_1, \dots, s_{2n} \in \{0,1\}} A^{(1)}_{s_1} A^{(2)}_{s_2} \dots A^{(2n)}_{s_{2n}} \ket{s_1, \dots, s_{2n}},
\end{equation}
where the bulk tensors $A^{(m)}_{s_m}$ are complex matrices of maximal dimension $\chi \times \chi$, and the boundary tensors $A^{(1)}_{s_1}$ and $A^{(2n)}_{s_{2n}}$ are strictly row and column vectors, respectively. The parameter $\chi$, denoted as the bond dimension, restricts the allowed entanglement entropy within the simulated network. In the context of classical fields, this entropy correlates directly with the complexity of spatial correlations and high-frequency modes generated by scattering interfaces.

\subsection{Shift Operators as Matrix Product Operators}

The iteration mapping requires applying the linear operators to the state vector. These are constructed as Matrix Product Ope-rators (MPOs). The elemental building blocks are the periodic shift operators $\hat{S}_x^{\pm}$ and $\hat{S}_z^{\pm}$  shown in \cite{kazeev2012low}, which translate the computational basis states along the respective coordinate axes.

The periodic shift operators are the exact mathematical representations of the quantum ripple-carry adders encoded within the MPO formalism. Let the spatial coordinate be mapped to the computational basis state $\ket{x} = \ket{x_{n-1} \dots x_1 x_0}$. The forward spatial shift executes the cyclic translation $\hat{S}^+ \ket{x} = \ket{(x+1) \bmod 2^n}$.

This transformation requires a bond dimension of exactly $\chi = 2$. The virtual bond index $\alpha_m \in \{0, 1\}$ acts as a deterministic carry bit propagating from the least significant spatial scale ($m=0$) to the most significant ($m=n-1$). We define the local transition operators on the $m$-th two-level subsystem as the identity $\hat{I}$, the raising operator $\hat{\sigma}^+ = \ket{1}\bra{0}$, and the lowering operator $\hat{\sigma}^- = \ket{0}\bra{1}$.

The bulk core tensor $W^{[m]}$ is formulated as a $2 \times 2$ matrix of these physical operators, where the row index corresponds to the incoming carry $\alpha_{m-1}$ and the column index corresponds to the outgoing carry $\alpha_m$:
\begin{equation}
W^{[m]} = \begin{pmatrix} \hat{I} & 0 \\ \hat{\sigma}^+ & \hat{\sigma}^- \end{pmatrix}
\end{equation}

To initiate the increment operation at the least significant bit, a carry of $1$ must be unconditionally injected. The boundary tensor $W^{[0]}$ is therefore restricted to the second row of the bulk matrix:
\begin{equation}
W^{[0]} = \begin{pmatrix} \hat{\sigma}^+ & \hat{\sigma}^- \end{pmatrix}
\end{equation}

The terminating tensor $W^{[n-1]}$ is constructed from the first column of the bulk matrix:
\begin{equation}
W^{[n-1]} = \begin{pmatrix} \hat{I} \\ \hat{\sigma}^+ \end{pmatrix}
\end{equation}

The complete forward shift MPO is the ordered contraction of these tensors across the virtual indices:
\begin{equation}
\hat{S}^+ = \sum_{\alpha} W^{[0]}_{\alpha_1} \otimes W^{[1]}_{\alpha_1, \alpha_2} \otimes \dots \otimes W^{[n-1]}_{\alpha_{n-1}}
\end{equation}

The backward shift operator $\hat{S}^-$ performs the translation $\ket{(x-1) \bmod 2^n}$. Because translation is unitary, the backward shift is the exact Hermitian adjoint of the forward operator, $(\hat{S}^+)^{\dagger}$. The corresponding bulk tensor requires taking the conjugate transpose of the physical operators while preserving the directional carry flow:
\begin{equation}
V^{[m]} = \begin{pmatrix} \hat{I} & 0 \\ \hat{\sigma}^- & \hat{\sigma}^+ \end{pmatrix}
\end{equation}

For the two-dimensional computational grid, these 1D MPOs are extended to $\hat{S}_x^{\pm}$ and $\hat{S}_z^{\pm}$ by taking the Kronecker product with the identity operator spanning the orthogonal spatial register:
\begin{equation}
\hat{S}_x^{\pm} = \hat{S}^{\pm} \otimes \hat{I}_{2^n}
\end{equation}
\begin{equation}
\hat{S}_z^{\pm} = \hat{I}_{2^n} \otimes \hat{S}^{\pm}
\end{equation}

\subsection{Matrix Product Operators and Evolution}

The spatial finite difference stencil for the Laplacian operator is formulated as an MPO sum:
\begin{equation}
    \Delta h^2 \hat{\nabla}_d^2 \approx \hat{S}_x^+ + \hat{S}_x^- + \hat{S}_z^+ + \hat{S}_z^- - 4\hat{I}.
\end{equation}
The velocity field $c_{i,j}^2$, which defines the background geological medium and high-contrast anomalies such as salt bodies, is encoded as a diagonal MPO, $\hat{C}^2$ as shown in \cite{oseledets2010tt}.

The discrete time evolution of the MPS is therefore given by the exact tensor algebraic equivalent of the LCU update equation:
\begin{align}
    \ket{\psi^{k+1}} &= 2\ket{\psi^k} - \ket{\psi^{k-1}}\nonumber \\  &+ \lambda^2 \hat{C}^2 \left( \hat{S}_x^+ + \hat{S}_x^- + \hat{S}_z^+ + \hat{S}_z^- - 4\hat{I} \right) \ket{\psi^k}.
\end{align}
Compare this MPS solver of the wave equation to the corresponding FDTD solver in equation (2).

Because the addition of two MPSs or the application of an MPO to an MPS multiplicatively increases the bond dimension, a compression step is strictly required after each operator application. The state undergoes a left-to-right orthogonalization sweep (QR decomposition) followed by a right-to-left SVD sweep. During the SVD step, singular values $\lambda_i$ below a specified cutoff ($\lambda_i < \epsilon \lambda_{\text{max}}$) are discarded, and the bond dimension is strictly truncated to $\max(\chi) = \chi_{\text{max}}$. This filtering mechanism ensures numerical tractability while preserving the dominant spatial structures of the propagating wavefronts traversing the subsurface.

\section{Results}
We apply the FTDT and the MPS solvers to the Marmousi data and demonstrate that 
the algorithm expressed in the MPS formulation generates valid 2D seismic wavefields.
We select a 256x256 grid subset of the Marmousi model and place an impulsive source of strength 1
in the center of the model at location of (1280 m, 1280 m). 
The spatial sampling $\Delta x=\Delta z$ is 10 m and the temporal stepping increment $\Delta t$ is 1 ms.  
The bond dimension is set to $\chi_{\text{max}}=128$ and the SVD cutoff value to $\lambda_{\text{max}}=10^{-7}$.
We propagate the wavefield from the center for 300 steps.

\setcounter{figure}{1}
\begin{figure}[htb] 
    \centering 
    \includegraphics[width=0.6\textwidth]{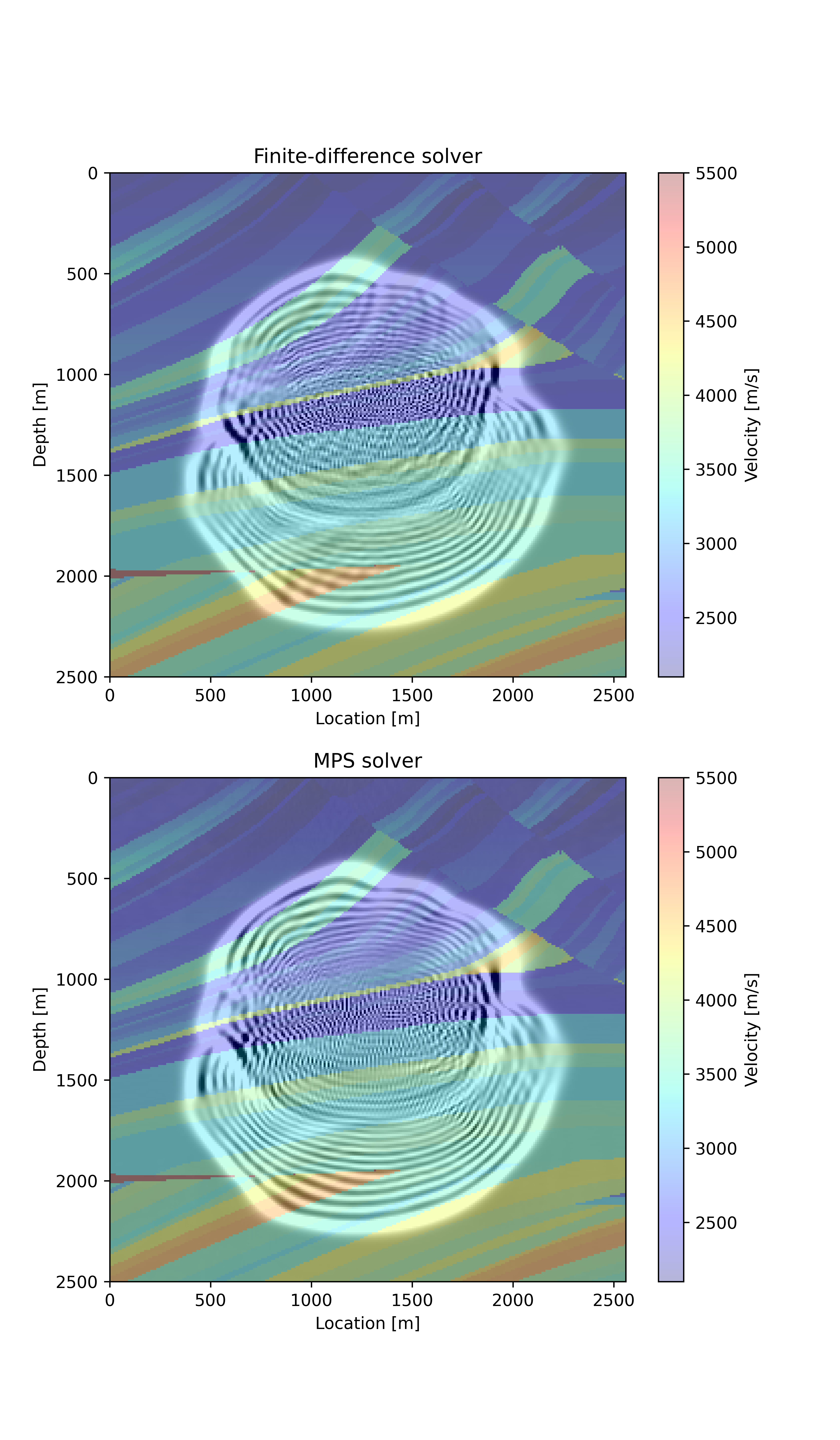} 
    \caption{Wavefields originating from the center of the model at location of (1280 m, 1280 m). Upper wavefield is calculated by an FDTD solver and the lower wavefield is by an MPS solver.}
    \label{fig:label}
\end{figure}

The results of the wavefield simulations are shown in Figure 2.
The top plot shows the wavefield generated by the FDTD solver over the Marmousi velocity model
and the bottom plot shows a similar arrangement for the wavefield generated by the MPS solver.
The MPS wavefield exhibits a close similarity to the FDTD waavefield, although some differences are visible upon closer inspection.
We attribute these discrepancies to the parameterization of the MPS solver and are conducting further analysis to better evaluate its potential capabilities.

\section{Discussion}
We have demonstrated that the MPS representation can produce 2D seismic wavefields on the relevant Marmousi model.
We are confident that 3D seismic wavefields can also be generated by the MPS solvers.

Table 1 compares the expressions describing the discretized wave equation for FDTD formulation, quantum encoding 
and MPS formulation from equations (2), (7) and (17), respectively.

\vspace{1cm}
\begin{table}[h!] 
    \centering 
    \label{tab:t1} 
    \vspace{1cm}
    \renewcommand{\arraystretch}{1.8}
    \begin{tabular}{|l|c|} 
        \hline 
        \textbf{Method} & \textbf{Equation} \\ 
        \hline 
        FDTD & $u_{i,j}^{k+1} = 2u_{i,j}^k - u_{i,j}^{k-1} + $ \\ 
             & $+ \lambda^2 c_{i,j}^2 \left( u_{i+1,j}^k + u_{i-1,j}^k + u_{i,j+1}^k + u_{i,j-1}^k - 4u_{i,j}^k \right)$ \\
        \hline
        Quantum & $\ket{\psi^{k+1}} = [2\hat{I} + \lambda^2 \hat{C}^2 (\hat{S}_x^+ \otimes \hat{I}_z + \hat{S}_x^- \otimes \hat{I}_z + $ \\
                & $+\hat{I}_x \otimes \hat{S}_z^+ + \hat{I}_x \otimes \hat{S}_z^- - 4\hat{I} )]\; \ket{\psi^k} - \ket{\psi^{k-1}}$  \\
        \hline
        MPS & $\ket{\psi^{k+1}} = 2\ket{\psi^k} - \ket{\psi^{k-1}}$ \\ 
            & $+ \lambda^2 \hat{C}^2 \left( \hat{S}_x^+ + \hat{S}_x^- + \hat{S}_z^+ + \hat{S}_z^- - 4\hat{I} \right) \ket{\psi^k}$  \\
        \hline 
    \end{tabular}
    \caption{Comparison of the wave equation representations} 
\end{table} 


Despite diverse formulations, all these expressions demonstrate
that the next wavefield is constructed from the current and previous wavefields. 
The next-step wavefield is computed via a linear combination of the current wavefield at local and neighboring grid points.
These methods differ in their representations of the 2D grid and the representations of the wavefield
values on these grids.
For the FDTD solver, the grids and the values are stored in the RAM of the hardware and values of neighboring values are accessed by the CPU via methods that favors fast streaming memory.
For the quantum solver, the grids and the neighboring values are encoded in entangled qubits and advanced using quantum gates.
The MPS solver uses the same classical hardware system as the FDTD solver, but grids and the wavefields are represented 
in a way that mimics the quantum system, using tensor networks. 
For MPS solvers, accessing the neighboring values are achieved using shift operators and organized into matrix product operators. 

Our limited experience in implementing MPS solvers for wavefield simulation so far indicates that these methods provide
an intriguing possibility computing and scaling wavefield propagation. 

\bibliographystyle{unsrt}  
\bibliography{references} 

\end{document}